\documentclass{article}%
\usepackage{amsfonts}
\usepackage{amsmath}
\usepackage{amssymb}
\usepackage{graphicx}%
\begin{document}

\title{Can Orbital Clustering of KBOs in the Ecliptic be due to the Solar Toroidal
Field Generated Spacetime Dragging?}
\author{Babur M. Mirza\\Department of Mathematics, \\Quaid-i-Azam University, Islamabad. 45320. PK.}
\maketitle

\begin{abstract}
The Kuiper belt objects (KBOs) exhibit an orbital clustering of the outer
planets lying at perihelion distances larger than Neptune and semimajor axes
greater than 150 AU from the Sun. This implies a hitherto unknown dynamical
mechanism to counter randomizing of the orbital elements caused by the giant
solar system planets. Using the toroidal field induced frame-dragging we
deduce here the observed range of the Kuiper belt region, the semi-major axis
of Sedna like objects in the Kuiper belt, as well as the orbital clustering of
the KBOs in the ecliptic, without assuming dynamical effects induced by
trans-Neptunian-objects (TNOs). We also calculate the orbital precession rates
for the inner planets and show their correspondence, within the range of
observational accuracy, with recent planetary ephemerides.

\end{abstract}

\section{Introduction}

The Kuiper belt objects (KBOs) span a broad range of perihelion distances,
ranging approximately from 35 AU to 80 AU. While some KBOs interact much more
strongly with Neptune, distant KBOs exhibit relatively stable orbits not
affected by Neptune's gravity. In the Kuiper belt, weak clustering of objects
undergo a rapid dynamical chaos which leaves little orbital structure in the
distant belt region (Batygin \& Brown 2016; Marcos \& Marcos 2014). The more
stable orbits, however, may still disperse due to the precession induced by
giant planets, including Jupiter, Saturn, Uranus, and Neptune. Numerical
simulations (Madigan \& McCourt 2015; Trujillo \& Sheppard 2014; Batygin,
Brown, \& Fraser 2011; Levison et al. 2008) of the orbital dynamics of the
distant KBOs show Sedna's orbital precession by 0.15 deg/Myrs, and 2014 SR349
precession at 0.8 deg/Myr. This range of precession rates leads to a far too
shorter time for the KBOs' orbits to disband, in approximately 100Myr. The
observed alignment of the orbital semi-major axis at a distance nearly 250 AU
implies a hitherto unknown dynamical mechanism, other than the secular
perturbation induced by the major planets inside the solar system.

In orbital dynamics of the KBOs, general relativistic effects, such as the
Lense-Thirring frame dragging of spacetime (Lense \& Thirring 1918; Rindler
1997; Tartaglia 2002; Iorio 2010), can be in general ruled out owing to their
extremely small magnitude. On the other hand, for inner planets, it has been
shown (Iorio 2018; Iorio 2012) that the Lense-Thirring frame dragging effect
has measurable magnitude which lie at the same or above the observational
error. It has also been used in detailed modelling of the planetary dynamics
in the solar system \footnote{For recent overviews of GR and the challenges
faced, see, e,g., (Iorio, 2005; Debono \& Smoot 2016) and references
therein.}. The on-going efforts to measure the relativistic frame-dragging
effect near the Earth, include high precision artificial satellite
measurements (Lucchesi et.al. 2019; Renzetti 2013). On the other hand long
range gravitational effects, such as those exhibited in the orbital dynamics
of the KBOs, cannot be accounted for by typical general relativistic effects.
In this case the dragging induced by the giant planets, and even that due to
the Sun, is negligible. Moreover, the metastable/stable orbital dynamics of
KBOs imply a cut-off region, where as the effective potential due to the
Lense-Thirring frame dragging decays as inverse cube of the orbital distance.
Similarly, the modification induced by the solar Lense-Thirring frame dragging
can be ruled out as significant in the dynamics of KBOs in view of the
clustering of the KBOs in the ecliptic. Within the Newtonian dynamics, the
clustering of KBOs orbits, can however be modelled under the postulate of a
trans-Neptunian planet (Batygin \& Brown 2016a; Batygin \& Brown 2016b; Brown
\& Batygin 2016; Batygin et.al. 2019; Iorio 2017). Notably, in these models,
the nineth planet (Telisto) causes secular correction to the Hamiltonian which
has an additive harmonic form.

Although typical general relativistic effects remain small in the case of
dynamics around normal stars, in compact stars spacetime effects play a key
role in various energy generation mechanisms. For example, in these systems
the wrapping of spacetime, affected by the extremely high magnetic field
energy density, plays an important role in accretion dynamics (Mereghetti
2008; Thompson \& Duncan 1995, Harding \& Lai \ 2006; Makishima\textit{\ }%
2014; Ciolfi, 2014; Muhlberger \textit{et al.}, 2014; Mitchell \textit{et
al.}, 2015). The relativistic field amplification in these stars is up to the
order of $10^{12}G$, where as in magnetars the surface currents indicate
fields as high as $10^{15}G$ (Tiengo \textit{et al.} 2013). In such cases the
high energy density of the star is modified by the spacetime wrapping. \ This
results in various observable phenomena, such as dragging induced surface
currents and high temperature gradients in the stellar atmosphere. Moreover,
this indicates that coupling of magnetic field to the background spacetime can
be a far more effective mechanism in the stellar dynamics than spacetime
induced effects only. Here the magnetic field can not only affect accretion
around the star via the Lorentz force, but can also induce modifications in
the spacetime itself. The magnetic field energy density acts here as a direct
measure of the modification induced in the background spacetime, since it
contributes to the total energy density of the system.

The discovery of the toroidal magnetic field in the Sun and Sun-like stars
implies that normal stars also posses analogous conditions, although with
relatively weaker fields. In these cases, the toroidal field is largely
generated at the stellar surface, and is sustained by differential rotation
inside the star (Petit \textit{et al.} 2008). Comparable in magnitude to the
poloidal field, the toroidal field ranges from $1G$ to $10^{3}G$ during the
solar cycle. As in the case of dense stars, the magnetic field energy density
can cause modification in the spacetime structure around a normal star like
the Sun. For the Sun, the magnetic field directly affects charge accretion via
Lorentz force up to the magnetosphere, and has negligible long range effect on
dynamics in the solar system. However, outside the solar surface not only the
spacetime curvature but also spacetime frame-dragging has effects manifest in
the orbital dynamics of the inner planets, such as the relativistic perihelion
advance of Mercury. The coupling of magnetic field energy density to the
background spacetime can be a more effective means of modifying the orbital
dynamics since it can extend spacetime dragging, analogous to the toroidal
field energy induced dragging in compact gravitational sources. In these
cases, the magnetic field energy density acts as an additional, locally
distributed pressure, causing the spacetime dragging effects to amplify. In
fast rotating dense stars, like the neutron stars, spacetime wrapping is
closely bound to the surface of the star, causing the observed surface
currents in these objects. In comparison, for Sun-like gravitational systems,
the relatively weaker wrapping can extends much farther in space. The coupling
of spacetime wrapping can be viewed here as an increase in the magnetic field
lines (per unit volume) in vicinity of the rotating gravitational source.
Conversely, the magnetic field energy density causes local stresses in the
spacetime structure, thus inducing modifications in the gravitational field of
the star. The magnetic field energy density corresponds here directly to the
enhancement induced in the spacetime, particularly in the case of the toroidal
field coupling around the star. The effect can be significantly high in the
equatorial plane of the star where the dragging is maximum, and can cause
orbital recession. The coupled field dragging can be more effective in orbital
precession than the Lense-Thirring frame-dragging, since it has larger local
magnitude and a longer range. In observational tests, the effect of such a
long-range frame dragging may therefore be detectable both in the orbital
precession of the inner planets as well as in the orbital dynamics of the
outer solar system. As shown below, the coupling has the form of a secular
perturbation, as postulated for Telisto, and can induce long term cyclic
changes in the KBOs' orbits. The spacetime vorticity here counterbalances the
gravitational attraction by providing the outward velocity drag, extending at
very large distances.

Here we show that the solar toroidal field modification induced in the
spacetime frame-dragging exhibits observable secular dynamical trends in the
orbital dynamics around the Sun. We thus derive the anomalous shifts in the
inner planetary orbits in the solar system and, as a long-range dragging
effect, deduce the distance of the Kuiper belt region along with the high
eccentricity orbits of the KBOs, as well as their observed clustering in the
solar ecliptic. In this we do not assume the nineth planet hypothesis, or the
existence of extra-solar effects such as the galactic dark matter. The
coupling of toroidal field to the spacetime dragging is shown here to be
derivable from General Relativistic (GR) Maxwell equations in the next
section. Also, it can be shown generally that the modification in the dragging
of spacetime is due to the axial symmetry, hence valid for any rotating
gravitational source in stable equilibrium, such as that with a Kerr
background spacetime. We then deduce the toroidal magnetic field energy
density as a function of the Lense-Thirring frequency, and calculate the
energy density of the toroidal field. The modified potential is then used to
derive the orbital parameters of the KBOs, including the semi-major axis, the
drag velocity, radius of the Kuiper belt, and also radius of the solar system.
We moreover show that the perturbations induced in the inner planetary orbits
correspond to the observed anomalous extraperihelion precessions in inner
planetary orbits in the solar system.

\section{Toroidal Solar Field and the Spacetime Frame-Dragging}

The toroidal field induced spacetime dragging can be derived from coupling of
the magnetic field to the gravitational field of a massive source. For this we
consider the Maxwell equations for the electromagnetic field in a curved
spacetime around a rotating mass (Mirza 2017; P\'{e}tri 2013; Mirza 2007; Oron
2002). In this case the background spacetime is axially symmetric and, for
slowly rotating objects, is given by the linearized Kerr metric,
\begin{equation}
ds^{2}=-e^{2\Phi(r)}dt^{2}+e^{-2\Phi(r)}dr^{2}+d\Omega^{2}-2\omega(r)r^{2}%
\sin^{2}\theta d\varphi dt, \tag{1}%
\end{equation}
where $d\Omega^{2}=r^{2}d\theta^{2}+r^{2}\sin^{2}\theta d\varphi^{2}$, and
$e^{2\Phi(r)}=(1-2M/r)$. Here $\omega(r)$ is the Lense-Thirring frame-dragging
frequency, equal to $2j/r^{3}$, where as $j$ is the angular momentum of the
star with mass $M$ . Also, the velocity $4$-vector compatible with the metric
is that of a co-moving observer, given by $u^{\alpha}=(e^{-\Phi(r)},0,0,\omega
e^{-\Phi(r)})$.

General Relativistic Maxwell equations in a curved spacetime are given by
(Landau \& Lifshitz 1980),
\begin{equation}
\qquad F_{\alpha\beta,\gamma}+F_{\beta\gamma,\alpha}+F_{\gamma\alpha,\beta}=0,
\tag{2}%
\end{equation}%
\begin{equation}
\left(  \sqrt{-g}F^{\alpha\beta}\right)  _{,\beta}=4\pi\sqrt{-g}J^{\alpha},
\tag{3}%
\end{equation}
where $g$ represents the determinant of the metric tensor $g_{\alpha\beta}$,
and $F_{\alpha\beta}$ is the electromagnetic field tensor. It can be shown
that the assumption of an everywhere finite $J^{\alpha}$ leads to the
condition that in a co-moving frame $E^{\alpha}=0=E_{\alpha}$ (Lichnerowicz
1967). Therefore, for the exterior region, the electromagnetic field tensor
takes the form $F_{\alpha\beta}=\sqrt{-g}\epsilon_{\alpha\beta\gamma\delta
}u^{\gamma}B^{\delta}$, and in the contravariant components $F^{\alpha\beta
}=-(-g)^{-1/2}\epsilon^{\alpha\beta\gamma\delta}u_{\gamma}B_{\delta}$, where
$\epsilon_{\alpha\beta\gamma\delta}$ is the four index Levi -Civita symbol.

Maxwell equations (2) and (3) can be solved to give the magnetic field as a
function of $(t,r,\theta,\varphi)$. The full set of Maxwell equations forms a
system of eight coupled partial differential equations (Mirza 2007). Under the
condition that $\omega\neq0$ and $B_{t}\neq0$, it can be verified that the
magnetic field around the star has the form,%
\begin{equation}
(B^{\alpha})=B_{0}\left(  0,0,\frac{A(\theta)\sin\theta\sin\xi}{u_{t}r^{2}%
\sin^{2}\theta},\frac{\cos\xi}{u_{t}r^{2}\sin^{2}\theta}\right)  , \tag{4}%
\end{equation}
where $B_{0}$ is a constant, $\xi=\varphi-\omega t$. Also,
\begin{equation}
A(\theta)=\int\frac{d\theta}{\sin\theta}. \tag{5}%
\end{equation}

\section{Dragging Effect in the Orbits of KBOs and Inner Planetary Orbits}

According to equations (4) and (5), the contribution of the poloidal field to
the total field energy density for any close loop for $\theta$ vanishes.
However, the toroidal field contributes to the energy density of the
surrounding field, for a rotating source, by
\begin{equation}
\epsilon_{\varphi}=\frac{B_{0}^{2}\cos^{2}\xi}{u_{t}^{2}r^{2}\sin^{2}\theta}.
\tag{6}%
\end{equation}
The field energy density is minimum in the equatorial plane ($\theta=\pi/2$)
and maximum at the poles. This is significant for the orbital clustering,
since the toroidal field energy density provides the outward velocity drag
around the Sun.

To calculate the velocity drag induced by the solar toroidal field, we
consider dragging effects in the equatorial plane $\theta=\pi/2$ of the Sun.
We therefore have for the solar toroidal field,%
\begin{equation}
B_{\varphi}=-\frac{B_{0}}{r\sqrt{1-\frac{R_{s}}{r}}}\cos(\varphi-\omega t),
\tag{7}%
\end{equation}
where $r>R_{s}$, and $R_{s}=2GM/c^{2}$ is the Schwarzschild radius.

For a slowly rotating star like the Sun, $\omega<<1$, and the cosine factor
depends on the choice of the coordinate $\varphi$, hence can be included in
the constant $B_{0}$. The radial drag therefore comes from the factor
$B_{0}/\left(  r\sqrt{1-R_{s}/r}\right)  $. Therefore, if $R$ denotes the
radius of the disc in the ecliptic (as the equatorial plane of the star),
containing a total mass $M$, then the total energy density of the field up to
the radius $R$ is given by,
\begin{equation}
\epsilon=\int_{0}^{R}4\pi r^{2}\epsilon_{\varphi}dr. \tag{8}%
\end{equation}
Substituting from equation (7) and integrating by parts we obtain,
\begin{equation}
\epsilon=4\pi B_{0}^{2}\left[  R+R_{s}\ln\left(  R-R_{s}\right)  \right]
\cos^{2}(\varphi-\omega(R)t),\text{ }R>R_{s}, \tag{9}%
\end{equation}
where $\omega^{2}$ and higher order powers have been neglected. We see that
this energy contributes to the total energy of the system, hence like the
gravitational potential energy, must be independent of the mass of the test
body. Also, the second term in equation (9) is very small as compared to the
first, and can be neglected. Therefore, the energy density due to the
toroidal-gravitational field coupling is,%
\begin{equation}
\epsilon=4\pi B_{0}^{2}R\cos^{2}(\varphi-\omega(R)t), \tag{10}%
\end{equation}
which depends on the frame-dragging frequency $\omega$ cyclically. In general,
for gravitationally bound systems with a power law potential function, the
virial theorem applies. For potential energy $V(R)$ and kinetic energy
$K=v_{d}^{2}/2$ (per unit mass), we thus have $2K=V=\epsilon$. Therefore, for
the radial velocity drag $v_{d}$, we have,%
\begin{equation}
v_{d}=B_{0}\sqrt{4\pi R}\cos(\varphi-\omega(R)t). \tag{11}%
\end{equation}
Equation (11) gives the outward (positive) drag velocity for a test body in
the gravitational field of the Sun, which is in addition to its orbital
velocity induced by the Newtonian gravitational potential.

Since $\omega<<1$, and $\varphi$ depends on the choice of the orientation of
the observer's coordinates, the energy density per unit mass due to
toroidal-gravitational field coupling is $\epsilon=4\pi B_{0}^{2}R$. Also, for
a test particle at a distance $R$, the gravitational potential due to the Sun
is $-GM/R$. Therefore, the total potential energy around the Sun, at a
distance $R$ is given by,%
\begin{equation}
\phi(R)=-\frac{GM}{R}+4\pi B_{0}^{2}R. \tag{12}%
\end{equation}
The additional contribution to the Newtonian gravitational potential here is
due to the modification in the gravitational field by the magnetic field
energy density. The effect of the new potential term correspond to a constant
acceleration depending on $B_{0}^{2}$. This corresponds to the extraperihelion
precessions, such as that observed in the case of the \textit{Pioneer}
anomaly. However, the acceleration depends on the magnetic field energy
density enclosed within an orbit, hence varies for different planetary orbits.
Also, the induced velocity drag depends on the orbital speed of the planet as
well (see also, Mirza 2019).

In Table 1 below, we give the shift $\Delta_{cal}=dv$ in the velocity of the
inner planetary orbits per planetary cycle, and compare it with the recent
observational data (Iorio 1019; Iorio 2015; and references therein). The
additional energy term above here corresponds to the dimensionless energy
$dE/2E=dv/v$, where $v$ is the orbital speed of the planet.%

\begin{tabular}
[c]{|l|l|l|l|l|l|}\hline
\textit{Planet} & $\theta(mas/cy)$ & $R(km)$ & $v(km/s)$ & $\Delta
_{obs}(km/cycle)$ & $\Delta_{cal}(km/cycle)$\\\hline
\textit{Mercury} & $-2.0\pm3.0$ & $58\times10^{6}$ & $47$ & $-8.8336$%
---$1.7667$ & $-3.5014$\\\hline
\textit{Venus} & $2.6\pm1.6$ & $108\times10^{6}$ & $47$ & $3.2898$---$13.8173$
& $5.7300$\\\hline
\textit{Earth} & $0.19\pm0.19$ & $150\times10^{6}$ & $30$ & $0$---$1.7362$ &
$1.4881$\\\hline
\textit{Mars} & $0.020\pm0.037$ & $228\times10^{6}$ & $24$ & $-0.1180$%
---$0.5000$ & $0.0895$\\\hline
\textit{Jupiter} & $58.2\pm28.3$ & $778\times10^{6}$ & $13$ & $708.6071$%
---$2049.9838$ & $2270.74$\\\hline
\textit{Saturn} & $0.15\pm0.65$ & $1.434\times10^{9}$ & $10$ & $-21.8403$%
---$34.9451$ & $4.5918$\\\hline
\end{tabular}

\textit{Table 1: The observed projected distance increase given by }%
$\Delta_{obs}=2\pi R\tan\left(  \theta\times10^{-3}/3600\right)  $\textit{, in
}$km$\textit{ per planetary cycle (with Earth year }$\approx3.1536\times
10^{7}s$)\textit{, compared with the calculated shift }$\Delta_{cal}$\textit{.
The observed planetary secular perihelion precessions }$\theta$\textit{ is in
milliarcseconds per century (}$mas/cy$\textit{), and the shifts are scaled per
century. Here }$R$\textit{ is the radial distance, and }$v$\textit{ is orbital
speed of the planet around the Sun.}

It is notable here that, whereas Mercury, Venus, and Saturn have no intrinsic
magnetic field, Earth, Mars, and Jupiter have locally generated equatorial
fields of magnitude approximately $0.3G$, $0.007G$, and $4G$, respectively.
This corresponds to a local velocity drag ($v_{local}\approx\sqrt
{B_{local}^{2}R_{planet}/2\pi}$), along the orbits of these planets, which is
given by $0.3390km/s$, $0.0194km/s$, and $51.7253km/s$, respectively. The
local amplification induced in the solar velocity drag is taken into account
in the calculated values of the shift $\Delta_{cal}$ for the planets with
significant magnetic field strengths. The calculated shift lies, within the
range of observation errors, in the same range (per planetary cycle) as observed.

In equation (12) the first term on the rhs. decreases inversely as the
distance, where as the second term increases linearly with increasing the disc
radius $R$. The Newtonian gravitational effects therefore dominate the
particle dynamics up to a distance $R_{0}$, such that $\phi(R_{0})=0$. This
corresponds to the minimum of the potential energy at $\phi(R_{0})=0$, which
gives for the effective range of the gravitational potential $R_{0}%
=\sqrt{GM/4\pi B_{0}^{2}}$. Putting the values of $M=M_{\odot}=1.989\times
10^{30}kg$, and $B_{0}=10G=10^{-3}T$ \ for normal solar activity period, we
obtain $R_{0}=31.82$ AU. This is comparable to the semi-major axis of Neptune
($\approx30.06$ AU). Therefore, in the solar system, the Newtonian
gravitational potential dominates the planetary dynamics up to Neptune (Fig.
1). The potential then changes sign and the repulsive potential due to the
toroidal field becomes effective. For Pluto (semi-major axis $\approx39.52$
AU), the toroidal field energy exceeds by a factor of $1.609\times10^{6}$,
whereas the drag velocity is approximately $0.22\times10^{-4}km/s$. This value
lies below the range of observational error as measured for the orbit of Pluto
in the recent flyby mission \textit{New Horizon }(Buie \& Folkner 2015). Also,
for Uranus ($19$ AU) and Neptune ($30$ AU), the velocity change is less than
$0.20\times10^{-4}km/s$ and $0.1\times10^{-4}km/s$, respectively, which also
lie below the range of observational error (Iorio \& Giudice 2006).

The drag velocity has the first zero at $\varphi-\omega(R)t=\pm\pi/2$, after
which the source effects of the dragging on the test particle vanish.
Neglecting the angular momentum due to the giant planets we have, at the
radius $R=R_{\max}$ in the ecliptic, $J=1.1\times10^{42}kgm^{2}/s$. Along the
axis $\varphi=0$, we have $R_{\max}^{3}=Jt/(\pi/2)$; which gives for an
orbital period around the Sun, $t=2.592\times10^{6}s$. Hence, for the extent
of the dragging effects we have the upper limit, approximately given by
$R_{\max}=10^{5}$ AU. This region correspond to the Kuiper belt in the solar
system. Here, as shown above, the repulsive toroidal field drag is greater
than the attractive Newtonian gravitational potential. Also, for distances
$R>R_{\max}$, only attractive gravitational potential remains effective, since
toroidal field effects reduce to zero at $R=R_{\max}$.

Using the field energy densities, we can also estimate the semi-major axis for
a planet whose orbit transverses the Kuiper belt region $R_{0}<R<R_{\max}$.
For an elliptical orbit, the semi-major axis $a$ of a test body is
proportional to the total energy $E$. At points of extrema, the velocity
vanishes (stationary points), therefore, the total energy equals the potential
energy at aphelion $r_{a}$. Denoting the semi-major axis under the
gravitational potential energy $E(r_{a})=GM(r_{a})/r_{a}$ by $a_{g}$, and by
$a_{\varphi}$ due to the toroidal field only, we have,%
\begin{equation}
\frac{a_{\varphi}}{a_{g}}=\frac{(4\pi B_{0}^{2}r_{a})V}{GM(r_{a})/r_{a}}.
\tag{13}%
\end{equation}
where $V=4\pi r_{a}^{3}/3$. Putting for the solar surface $r_{a}\approx
R_{\odot}$, we obtain the scaling relation $a_{\varphi}\approx(10^{15}m)a_{g}%
$. Equation (13) implies that the semi-major axis in an orbit around the Sun
increases by a factor of $10^{4}$ AU under the toroidal field. This
corresponds to the observed high eccentricity of KBOs. It therefore follows
that outside the disc, with radius of approximately $31.82$ AU, the outward
dragging causes velocity increase in the planetary orbits, giving the highly
eccentric periodic orbital motion observed for the KBOs.

\section{Conclusions and Summary}

In the above, we have shown that the coupled field dragging effect not only
induces orbital changes in the inner planetary orbits, but has long range
effects in the solar system, extending up to the Kuiper belt. For the KBOs,
modification in the spacetime dragging causes the following observable features.

(1) The clustering of the KBOs in the solar ecliptic is due to the minimum of
the outward spacetime drag which, according to equation (6), lies in the
equatorial plane of the Sun (see also, Fig. 2).

(2) The minimum of the potential energy function is used to determine the
distance up to which the Newtonian gravitational potential dominates the
planetary motion. This gives for the range of the inner solar system the
approximate distance up to Neptune. For distances larger than Neptune,
toroidal field dragging plays a more effective role in the orbital dynamics
than the Newtonian gravitational potential. This causes the formation of a
Kuiper belt-like region.

(3) In the Kuiper belt, the coupled outward drag has maximum magnitude in the
solar ecliptic, which allows KBOs to move through the ecliptic, with increased
velocity and high eccentricity, in orbits of larger semi-major axes.

Summarizing, the effects of the spacetime dragging have been calculated here
for the planetary orbits around the Sun. For the inner planets, the spacetime
dragging induces a cyclic shift in the orbits. It was found that the derived
magnitudes of the shift in the orbital velocity of the inner planets
correspond to the observed shifts, lying within the range of observational accuracy.

We have also shown here that the orbital dynamics of KBOs is dominated by the
toroidal field induced spacetime dragging. The coupling of the solar
gravitational field to the solar toroidal magnetic field thus modifies the
gravitational field of the Sun. This modification exceeds the Newtonian
gravitational potential at sufficiently large distances. In contrast with the
strong spacetime wrapping around rotating compact stars, the solar magnetic
field generated dragging extends much farther in space for normal stars. For
Sun-like stars, this is due to the weaker spacetime wrapping, since the lines
of force (per unit volume) are relatively less closely bound around the star.

\textbf{Acknowledgement}

I gratefully acknowledge the anonymous reviewer for suggestions regarding
recent planetary ephemerides.

\bigskip

Figure Captions:

Figure 1: The total potential energy function (12) per unit gravitational
potential energy (G.P.E.) of a planet. The gravitational potential dominates
approximately up to the orbit of Neptune. Here the corresponding Kuiper belt
region extends to the radius between $31.82$ AU to $10^{5}$AU, where the
toroidal potential is effectively higher than the Newtonian gravitational potential.

Figure 2: Orbital clustering in the ecliptic due to the local minima of the
toroidal potential in the plane $\theta=\pi/2$. The potential is maximum at
the poles, causing the KBOs to group towards the ecliptic plane, particularly
while crossing the Kuiper belt where the toroidal potential energy exceeds the
Newtonian gravitational potential.

\end{document}